# The building blocks behind the Electrohydrodynamics of non-polar 2D-inks

*Pedro C. Rijo[1,2] and Francisco J. Galindo-Rosales[2,3,]\**

[1]Transport Phenomena Research Center (CEFT), Mechanical Engineering Department, Faculty of Engineering, University of Porto, Rua Dr. Roberto Frias, 4200-465 Porto, Portugal.
[2]AliCE – Associate Laboratory in Chemical Engineering, Faculty of Engineering, University of Porto, Rua Dr. Roberto Frias, 4200-465 Porto, Portugal.
[3]Transport Phenomena Research Center (CEFT), Chemical Engineering Department, Faculty of Engineering, University of Porto, Rua Dr. Roberto Frias, 4200-465 Porto, Portugal.
\*E-mail: galindo@fe.up.pt

**Abstract**

This work provides a complete rheological characterization of 2D-inks in electric fields with different intensities and orientations to the imposed flow field. 2D nanomaterials used in this study are graphene nanoplatelets, hexagonal boron-nitride, and molybdenum disulfide. These materials with different electric properties are dispersed in a non-polar solvent (Toluene) with different concentrations of Ethyl Cellulose (EC), providing Newtonian or viscoelastic characteristics. Shear rheology tests show that the presence of nanoparticles barely changes the fluid behavior from the carrier fluid, and the application of an electric field perpendicular to the flow does not result in electrorheological behavior. However, extensional experiments, which mimic the actual EHD jet printing conditions, allowed the observation of the influence of both the particles and the electric field aligned on the filament thinning process. It was observed that the electric field generates vortices due to an electrophoretic effect in the carrier fluid when EC is present in the formulations, which has severe consequences on the stability of the liquid bridges, whereas it scarcely affects the shear viscosity; additionally, the kind of 2D nanoparticles modifies also the conductivity and permittivity of the solution, inducing Maxwell stresses that also make the filament more stable against surface tension.

# 1. Introduction

2D nanomaterials have massively attracted the scientific community's interest in recent years, mainly because their great aspect ratio and surface area allow improving some mechanical, electronic, and chemical properties that it is not possible to observe when particles of the same material are 3D in shape (graphene *vs* graphite).[1, 2] Further, surface modification and functionalization can easily change some of these properties.[1] All these unique features that make them especially useful for many applications, such as in biomedical diagnostic applications[3], photovoltaic cells[4], air filters[5], membranes for water treatment[6], or electrochemical energy storage technologies.[7] Additionally, the chemical structures and the optical properties of 2D nanomaterials have been found to be suitable to improve biosensor sensitivity and the detection limit that traditional sensors (such as electrical sensing) do not have.[8] Inkjet-printed devices incorporating 2D nanomaterials have been demonstrated[9]; however, this traditional printing process forces 2D-inks to match certain rheological properties and sometimes rheology modifiers are added to the ink's formulation, which in most cases ruins the final properties of the ink.[10] The typical formulation of 2D-ink consists in the dispersion of 2D nanomaterials in a polymer solution to avoid sedimentation of the particles during the manufacturing process, which exhibit viscoelasticity and behave as complex fluids. So, the determination of the full rheological properties of the 2D-inks is crucial for defining a constitutive equation and predicting their flow behavior in real processes[11, 12]; moreover, it is also possible to prevent the use of rheology modifiers and adapt the parameters of the printing processes to the rheology of the ink.[13] As happens for any complex fluids, most of the works reported in the literature deal with the rheological characterization of 2D nanoparticles under shear flow.[14-18] Results showed a clear the lubricating effect at low concentrations and that the Einstein equation and Hinch-Leal equation cannot predict viscosity as a function of particle concentration because both models do not have into consideration the interaction between the particles and the fact that the particles have oblate shapes[19]; however, for higher concentrations, suspensions follow the Bingham plastic fluid model.[19] Even though the extensional flow dominates in some parts of the printing processes[20], extensional rheometry is less developed than its shear counterpart due to intrinsic difficulties. However, the extensional rheometry helps to understand how the droplets breakup mechanism occurs (**Figure 1**) to avoid the appearance of defects on printed products. Few works report the flow behavior of inks under extensional flow[21]; moreover, the literature is even more scarcer for 2D-inks[22]; and unexisting when an external field is applied aligned to the direction of the extensional flow, which mimics the actual printing conditions. Rijo *et al.* [17] were pioneers in the extensional rheological characterization of this type of fluid to understand how the presence of 2D nanoparticles influences the relaxation time of the fluid and the formation of beads-on-a-string just before filament breakage. They observed that the presence of 2D particles limits the polymer chains to stretch which decreases the relaxation times of the inks and delays the formation of a perfect bead-on-a-string before the filament breakage.

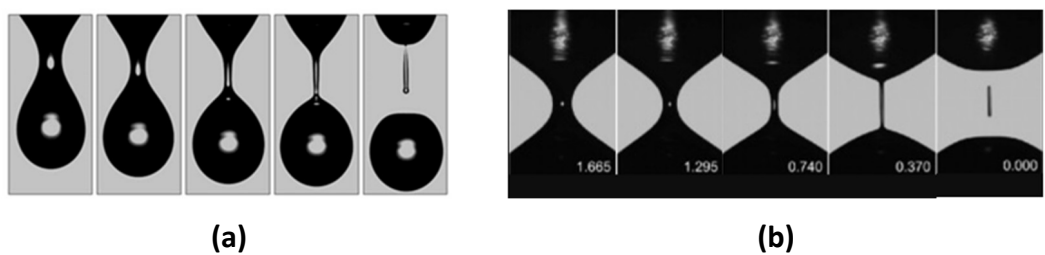

(a)  (b)

**Figure 1.** Sequence of images comparing **(a)** the drop formation and breakup of 70% glycerol/water mixture with **(b)** the filament thinning from CaBER of 75% glycerol/water mixture. Reprinted and adapted from[23, 24].

Among the plethora of techniques for printed electronics, jet printing-based manufacturing processes are very attractive because of the ability to generate very small-scale droplets, whereas inkjet printing using thermal or piezo-excitation represents a highly established and successful approach for flexible electronic manufacturing, relying on the localized delivery of materials of interest directly to substrates with high spatial control. However, conventional inkjet printing is considered a "push" printing process providing droplet sizes usually several times larger than the nozzle diameter, besides leading to various intrinsic problems of nozzle blockage, limited ink adaptability and a limited resolution of ~20 μm. In contrast to this process, the electrohydrodynamic (EHD) jet printing technique is a "pull" process as an extra electric field is introduced, which allows for increasing the resolution and throughput, representing an ideal alternative to conventional thermal and piezoelectric inkjet systems.[20] EHD jet printing is a noncontact printing technique that has gained much attention in recent years because it can induce various jetting modes by applying the inkjet printing process, which is a representative of "drop-on-demand" digital printing method, under an electric field.[25] This field deforms the meniscus of the droplet to a conical shape known as a Taylor cone, which can consequently emit a jet of ink droplets. This printing method has the advantage of forming sub-micrometer ink droplets and a highly stable undisrupted jet of ink under a uniform electric field. The quality of the products obtained from EHD jet printing depends on the rheological properties of the inks, working distance, the voltage applied to the system, etc.[25-27] However, the influence of the electric field on the rheological properties of the inks has been systematically neglected, especially in 2D inks. Again, the literature is scarce regarding electrorheological studies of particulate suspensions. Several authors[28-35] studied the influence of particle concentration, particle shape, size, electric conductivity, etc., on the electrorheological properties of suspensions containing 3D particles from micro to nanoscale or nanofibers. So far, Mrlik *et al.*[36], Yin *et al.*[37] and Lee *et al.*[38] studied the electrorheological effect of 2D particles dispersed in a Newtonian fluid. They found that the presence of 2D particles improves the dielectric properties since the polarization effect is the primary mechanism responsible for the appearance of ER effect. Moreover, they observed that the temperature and particle surface modification affect the ER effect.

To the best of the authors' knowledge, it has not yet been reported the electrorheological response of 2D nanoparticles dispersed in a viscoelastic fluid when the electric field is parallel to the flow direction. In this study, the empty gap in the literature is filled by evaluating the shear and extensional electrorheological properties of non-polar 2D-inks, loaded with nanoparticles of different electric conductivities, with the aim of understanding how the electric field affects their key rheological properties, i.e., shear viscosity and the relaxation time. This study will allow us to understand which are the building blocks behind the electrohydrodynamics of non-polar 2D-inks that will allow to improve the printing quality of 2D-inks by means of EHD jetting.

## 2. Materials and Methods
### 2.1. Materials

The nanomaterials and reagents used for the formulation of 2D-inks were ethyl cellulose (48% ethoxyl basis), toluene (purity > 99.9%), graphene nanoplatelets (GNP), hexagonal boron nitride (hBN) powder, and molybdenum disulfide ($MoS_2$) powder. Sigma-Aldrich supplied all materials except toluene and GNP, supplied by Carlos Erba Reagents and Graphenest, respectively. The viscoelastic fluid consists of the dissolution of ethyl cellulose in toluene for several concentrations. The concentrations used were 2.5% w/v (weight by volume) and 5% w/v. Pure toluene was considered the Newtonian reference fluid. The 2D nanomaterials were dispersed with a concentration of 0.2 mg/mL. The preparation protocol and the formulation of these 2D-inks are the same as those used in our previous work. [17]

## 2.2. Dielectric Properties and electrical conductivity

A Keysight E4980AL LCR meter was used to measure the dielectric properties of the fluids. The experimental tests were done at 20 °C and 1 V in an alternating current (AC) frequency window between 100 Hz and $10^5$ Hz. The liquid was confined in two horizontal stainless steel plates with 50 mm diameter and the distance between the plates was 0.5 mm. These tests allowed us to estimate the electrorheological potential of the fluids.

To measure the electrical conductivity of the fluids, a Tektronix/ Keithley Model 6482 dual-channel picoammeter was used with the same parallel plates and the same gap. Experiments were performed at 20 °C and imposing 10, 20 and 30 V in direct current (DC). Three independent runs were done with fresh samples and the measurements of voltage and current were taken 30 seconds after the application of the voltage.

## 2.3. Rheological Measurements

### 2.3.1. Shear Rheology

A controlled shear-stress rheometer (Anton Paar MCR301) was used to determine the shear viscosity of the suspensions when an electric field is applied (**Figure 2**). The electro-rheological device (ERD) consists of plate-plate geometry with a 50 mm diameter connected to a high voltage power supply (FUG HCL 14-12500). The distance between the plates was 0.1 mm and all tests were performed for a temperature of 20 °C. For each suspension, 5 independent runs were done to ensure good repeatability of the results.

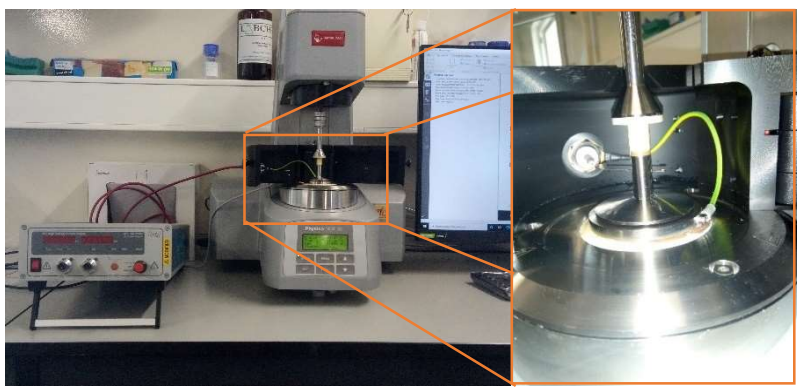

**Figure 2.** Experimental setup used to measure steady-shear viscosity and the electrorheological cell used to apply electric field to the fluids.

Steady-shear viscosity tests were performed for a range of shear rates of 1 to $10^5$ s$^{-1}$. It is important to highlight the fact that the shear flow and electric field are perpendicular to each other under this configuration (Supplementary Information, section SI1 for further information on the reliability of the shear electrorheological measurements).

*4.2.2. Extensional Rheology*

In this work a capillary breakup extensional rheometer (CaBER) is used to perform the extensional rheological characterization. The experimental setup used is depicted in **Figure 3a**. The evolution of the filament diameter was recorded using a high-speed camera (Photron FASTCAM mini UX100) coupled with a set of optical lenses (Optem Zoom 70XL) with a variable magnification from 1X to 5.5X. To visualize the filament shape, it was necessary to use a 60 mm Telecentric Backlight Illuminator (TECHSPEC), where a white light was supplied by a fibre optic cable connected to a light source (Leica EL6000). The image analysis was done using an in-house developed code in MATLAB, which determines the minimum filament radius, the extensional rate, and extensional viscosity. All tests without the application of the electric field use the 4 mm plate geometry provided by the manufacturer (**Figure 3b**). This rheometer is not prepared to perform tests with the application of an electric field. To overcome this problem, Sadek *et al.*[39] developed an electrorheological cell that can apply an electric field without damaging the rheometer and interfering with its normal functioning. The final prototype of the plates used here is shown in **Figure 3c,** the diameter of metal plate is also 4 mm and the high voltage power supply (LabSmith HVS448-3000) is used to applied an electric field to the fluid. It is important to highlight the fact that the extensional flow and electric field are parallel to each other under this configuration.

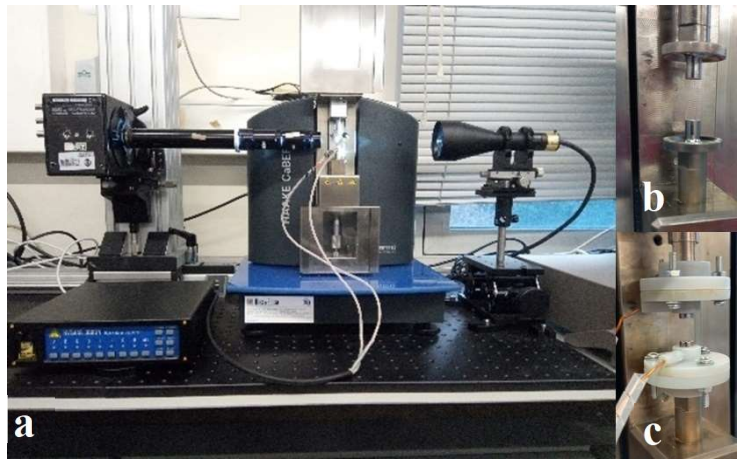

**Figure 3. (a)** Experimental setup used for extensional tests. **(b)** Standard 4 mm plate geometry used during the experiments without electric field. **(c)** Electrified 4 mm plate geometry used during the experiments when an electric field is applied.

According to the numerical results of Yao and McKinley [40] for filament stretching rheometry, the initial aspect ratio ($\Lambda_0$) would be optimal in the range of $0.5 \leq \Lambda_0 \leq 1$. As the 4 mm diameter plates are selected. $\Lambda_0$ is set at 0.5 due to the working fluid's low value of surface tension. The initial aspect ratio is the ratio between the initial gap between the plates ($h_0$) and the diameter of the plate, so the $h_0$ is fixed at 2 mm. This value ensures that the interfacial force from surface tension can support the liquid bridge against the sagging induced by the gravitational body forces [41]. Knowing the time evolution of the minimum radius ($R_{min}(t)$) and the extensional rate ($\dot{\varepsilon}$) during the lifetime of the filament, the extensional viscosity ($\eta_E$) can be determined as follows:

$$\eta_E = \frac{\sigma}{R_{min}(t)} \cdot \frac{1}{\dot{\varepsilon}} \tag{1}$$

$$\dot{\varepsilon} = -\frac{2}{R_m(t)} \cdot \frac{dR_m(t)}{dt} \tag{2}$$

where $\sigma$ represents the surface tension of the fluid. The slow retraction method (SRM) was used in order to minimize the inertia effects and allow the measurement of relaxation times with the help of a high-speed camera.[24]

## 3. Results and discussion

### 3.1. Dielectric properties and electrical conductivities

The interfacial polarization, also known as Maxwell-Wagner polarization, is the main mechanism responsible for the electrorheological effect of suspensions. According to Block *et al.* [42], the relaxation frequency of the ER fluid is proportional to the polarization rate, which must be in the range of 100-$10^5$ Hz for a good electrorheological effect. In accordance with Ikazaki *et al.*[43], the polarization rate maintains the chain-like structures formed by the particles dispersed in a carrier fluid under an applied electric field. Whether or not the chain-like structure is maintained under the flow conditions is determined by the relation between the rotational speed and the polarization rate of the dispersed particles. When the polarization rate is too low, it cannot create these structures, but the structures are formed if it is too high. They are easily rotated by the motion induced by the shear gradient. The induced dipoles in the particles will realign according to the electric field direction, promoting the repulsive forces between particles and not the attractive forces to sustain the particle structures.[43] Moreover, the difference in dielectric constants between 100 Hz and $10^5$ Hz is also a parameter to be considered to study the electrorheological effect. If the relaxation frequency is in the range of 100-$10^5$ Hz, the higher the difference of the dielectric constants, the greater the electrorheological effect will be.[43]

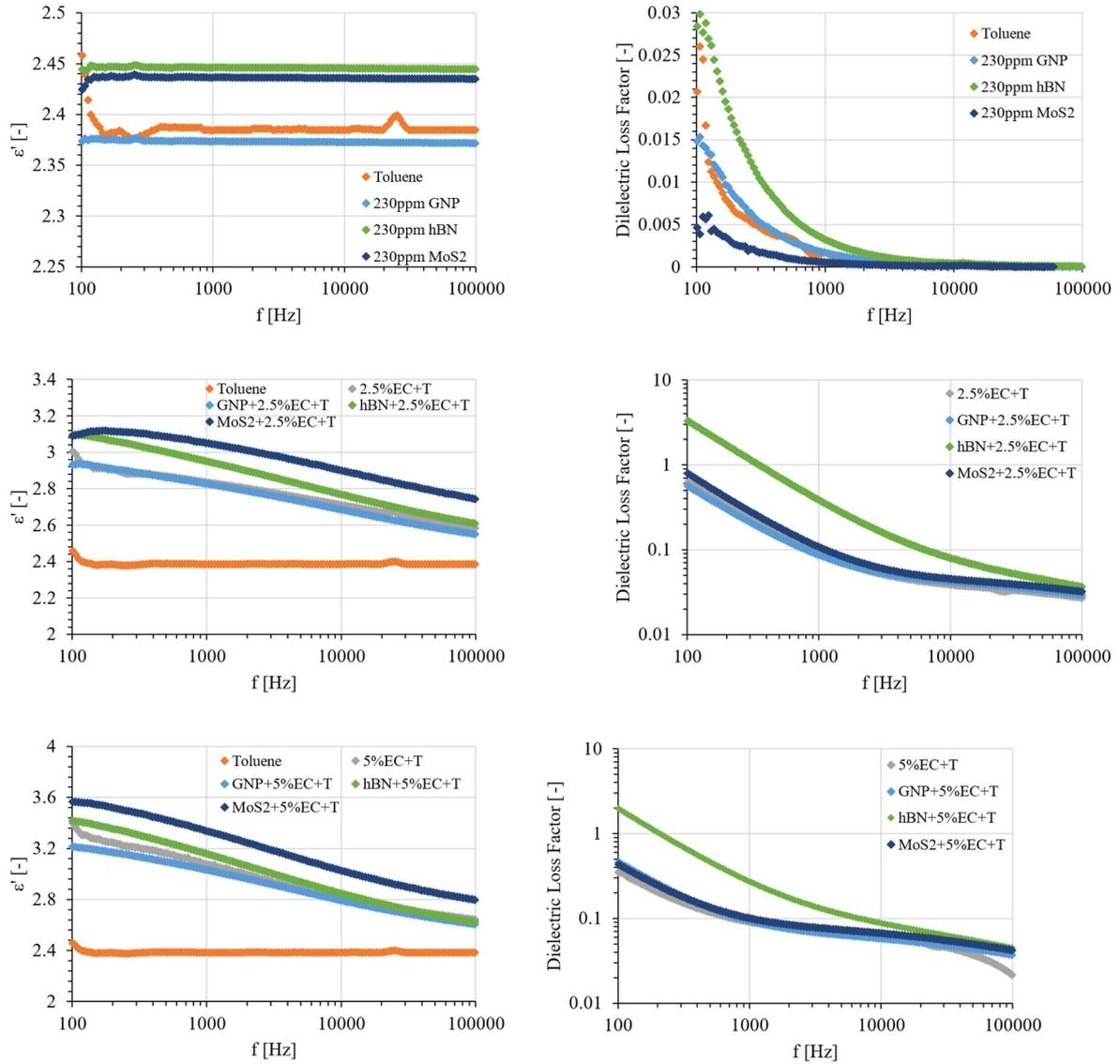

**Figure 4.** Dielectric constant (ε') and loss factor for all 2Dinks considered in this work.

**Figure 4** shows the dielectric constant and dielectric loss factor curves for all fluids studied in this work. The dielectric constant of toluene is practically independent of the frequency, and it is equal to 2.387 ± 0.013 which agrees with data found in the literature.[44] Regarding the polymeric solutions, the curves present in both figures follow the same trend verified by Nojiri and Okamoto [45] for the poly(vinyl acetate)-toluene solution. These authors also found that solution temperature, polymer concentration and polymer molecular weight influence the dielectric constant and relaxation frequency. When particles are added to the suspension, the relaxation frequency is found to be below 100 Hz. Thus, the polarization rate is too low to allow the formation of particle structures that promote the electrorheological effect. Furthermore, the difference in dielectric constants (Δε') between 100 Hz and $10^5$ Hz (**Table 1**) is lower than 0.80.

**Table 1.** Evolution of dielectric constant drop ($\Delta\varepsilon' = \varepsilon'_{10^2 Hz} - \varepsilon'_{10^5 Hz}$) for all fluids at 20 °C.

| Fluid: | Toluene | 2.5% w/v EC + Tol. | 5% w/v EC + Tol. |
|---|---|---|---|
| Particles | $\Delta\varepsilon'$ | $\Delta\varepsilon'$ | $\Delta\varepsilon'$ |
| No Particles | ≈ 0 | 0.421 | 0.756 |
| GNP | 0.002 | 0.386 | 0.611 |
| hBN | ≈ 0.0 | 0.490 | 0.796 |
| MoS$_2$ | 0.006 | 0.349 | 0.773 |

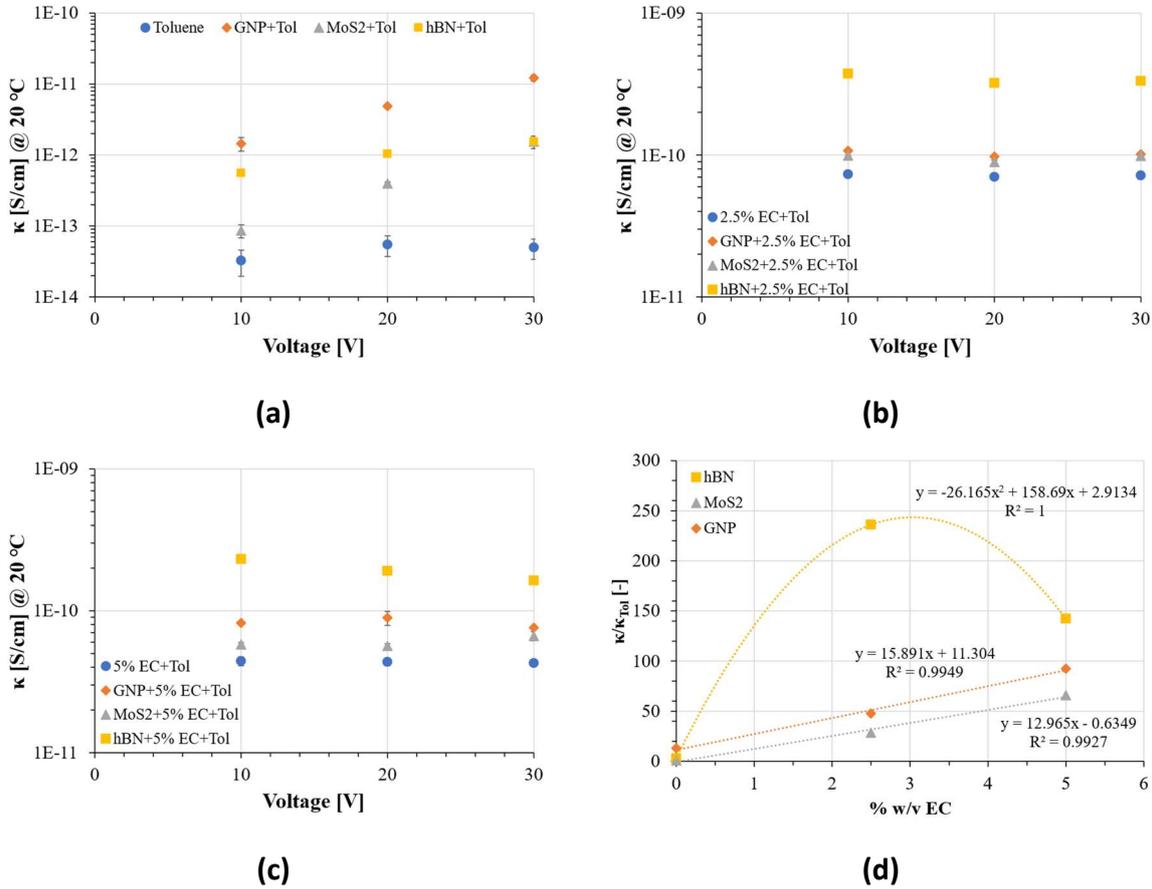

**Figure 5.** Electrical conductivity for all 2Dinks considered in this work: a) Toluene, b) 2.5% w/v EC+Tol. and c) 5% w/v EC+Tol. as carrier fluids. d) relative conductivity of the 2D-inks regarding the conductivity of the Toluene.

**Figure 5a-c** shows the electrical conductivity of the 12 working fluids. It can be observed that the electrical conductivity is very small for the inks without EC, and it grows with the imposed

voltage, as expected.[46] The presence of EC increases the conductivity (above two orders of magnitude) and stabilizes its value for growing voltages. It can be observed in **Figure 5d** that the conductivity of the 2D-ink increases linearly with the concentration of EC for GNP and $MoS_2$ nanoparticles, whereas it follows a parabolic dependence for the hBN, being maximum near 2.5%. Interestingly, the conductivity of the ink is not the largest for the most conductive nanoparticle (GNP), but for the least (hBN). This latter result deserves further research, which lies out of the scope of this work.

**3.2. Shear electrorheological characterization**

Without the application of an external electric field, Toluene exhibits a Newtonian behavior; moreover, the application of an external electric field perpendicular to the flow direction does not affect its rheological behavior, as expected from its dielectric properties (section 3.1). The presence of the 2D nanoparticles nor the application of an external electric field deviated from the viscosity curve of Toluene (**Figure 6** and **Figure SI2**), due to the low mass fraction of particles (230 ppm). In contrast to the electrorheological effect observed for mass fractions lower than 0.2% w/w for graphene oxide dispersed in silicone oil. [47] It is noticeable that for GNP suspensions, the electric field strengths higher than 1.0 kV/mm are truly applied only for shear rates higher than $10^3$ $s^{-1}$, before that the intensity of the electric field increases as the shear rate increases (**Figure 7a**). Although toluene has a low surface tension and dielectric constant, facilitating the exfoliation and dispersion of graphene sheets, toluene is not a potent stabilizing agent due to its low viscosity, which allows nanoparticle aggregation. When an electric field is applied, the nanoparticle migration to the positive electrode occurs, and the particles are randomly deposited in the upper plate, which can have small number of contact points between the two plates, as schematically represented in **Figure 7c**; this phase separation (electromigration) phenomenon has been reported in the literature as a source of negative electrorheological effect, where the viscosity or rheological properties decrease in the presence of an applied electric field.[48] Graphene aggregation occurs through van der Waals forces, which can be high enough to avoid particle disaggregation at low shear rates. Applying of an electric field can also increase the strength of the van der Waals force.[49] When the shear rate increases, a partial disaggregation of particles can occur, and they will be deposited on the positive electrode in a compact form, which reduces the number of contact points and allows an increase in the voltage supplied by the source since the electric conductivity of the 2D-ink reduces. For hBN and $MoS_2$ particles, the electric field strength applied to the fluid is respected for the entire range of the applied shear rate, due to their lower conductivity regarding the one of GNP.[50, 51]

When the carrier fluid contains ethyl cellulose, its viscosity increases and reduces the Brownian motion that prevents the nanoparticle aggregation, and reliable results can be obtained at lower shear rates with the ERD cell than for pure Toluene. At 2.5%w/v EC, the application of an external electric field promotes a further slight increment in shear stress for shear rates between 100 s$^{-1}$ and 7000 s$^{-1}$; however, data below 100 s$^{-1}$ is unreliable due to the presence of the wire, and for shear rates higher than 7000 s$^{-1}$, the measured shear stress values for each electric field approximate the shear stress values measured without electric field (**Figure 6 and Figure SI3**). This latter behavior is due to the competition between the electric field and the flow field,[27] and for high shear rates, the flow field overcomes the electric field and controls the process. This apparent increase of the shear stress is not due to the reorientation of the nanoparticles toward the electric field but rather to the vortex formation inside of the fluid, as shown in **Figure 7b** and better visualized in the **movies SI1** and **SI2** present in the supporting information. According to Barrero *et al.* [52], the formation of vortices inside of the fluid can be derived by (i) the tangential electrical stresses acting on the liquid-gas interface and (ii) the flow rate injected to the electrified needle. In this work, the second condition is not verified since the fluid is sandwiched between the electrodes and the volume of fluid is kept constant. Moreover, the intensity of the vortices depends on the electrical conductivity and viscosity of the fluid and the vortices are more intense when the fluid exhibits both low electrical conductivity and low viscosity. This phenomenon is frequently observed in Taylor's cones in the presence of electric field and in static condition (flow rate is null).

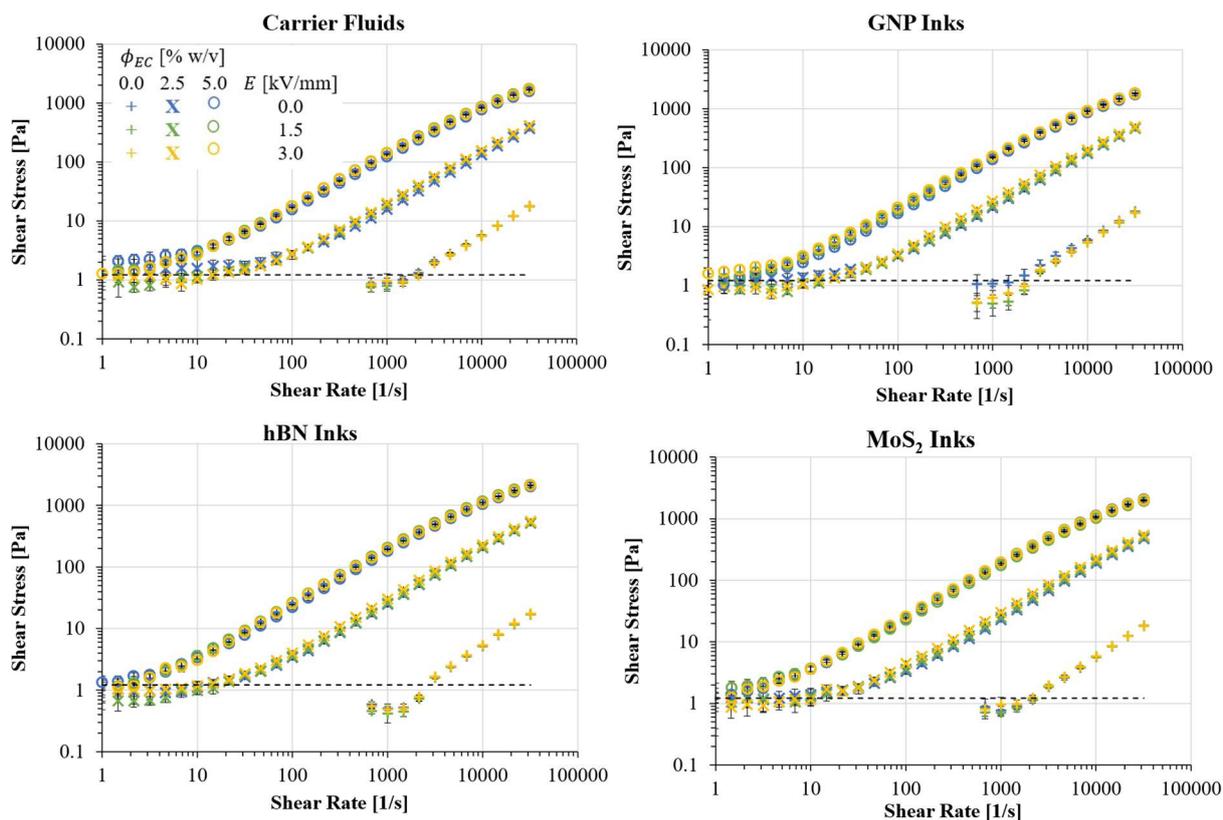

**Figure 6.** Shear electrorheological behaviour of 2D-inks.

When an electric field is applied, the presence of the vortices inside the fluid results in a very similar behavior to that observed when secondary flows induced by inertia/centrifugal forces are present.[53] This phenomenon promotes an increase in the shear stress, which in turn corresponds to an increase in fluid viscosity. Although the vortices are present as long as the electric field is active, their influence on the flow curve will be sensed more intensively at low shear rates. In contrast, it will be negligible at high shear rates, providing the illusion of an inexistent yield stress behavior. When polymer solution is at 5% w/v of ethyl cellulose, **Figure 6** shows a further increase of the shear stress at low shear rates when the electric field strength increases, and there is a saturation in the shear stress curves for electric fields higher than 2.0 kV/mm. This saturation effect was also reported in the work of Pereira [54], and it is due to a higher number of interactions between polymer molecules when polymer concentration increases, which difficult the orientation of the molecules in the direction of the electric field and weaken the vortex velocity. At such a low concentration, the electrorheological effect induced by the 2D nanoparticles (**Figure 6** and **Figure SI 4**) will barely affect the dynamic of the shear flow with a perpendicular electric field since the particles migrate to the positive electrode and only the vortices induced by the EC affect the rheological properties in the presence of electric field.

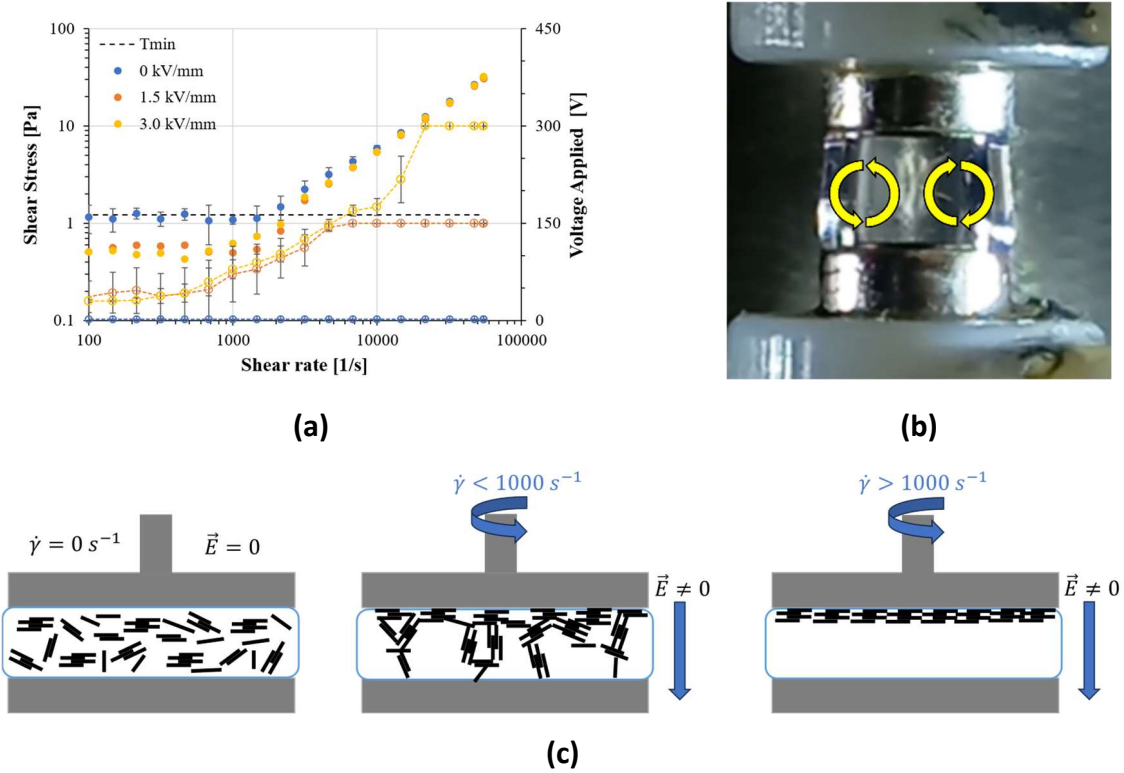

(a) (b)

(c)

**Figure 7. (a)** Real voltage applied (dashed lines) to GNP dispersed in toluene during the steady-shear test. **(b)** Vortices orientation in 2.5% w/v of ethyl cellulose polymer solution when an electric field of 1.5 kV/mm is applied (plates of 4mm in diameter separated by a gap of 2 mm). **(c)** schematic representation of particle migration under the influence of an applied electric field and a shear-rate.

## 3.3. Extensional electrorheological characterization

In this section it will be discussed the extensional behavior of the different 2D-inks considered in this study. The graphs below were obtained from the analysis of the images recorded by means of a high-speed camera (**Figures SI6-SI16**), where it becomes evident the influence of the external electric field when it is aligned with the direction of the flow. In CaBER experiments, the electric field strength decreases as the distance between the plates increases and the electric field strength decreases as follows:

$$E(t) = \frac{V_0}{H_0 + v_{pt}t} \tag{3}$$

where $V_0$ is the voltage applied to the fluid, $H_0$ is the initial gap between the plates, $v_{pt}$ is the velocity of the upper plate and $t$ is the time. Based on this, the electric field strength shown in this section and the corresponding support information refers to the initial electric field strength applied to the fluid.

**Figure 8** shows the time evolution of the minimum filament radius of toluene for various electric field strengths. In these experimental data, the Ohnesorge (Oh) number is lower than 0.2077 and, consequently, the decrease of the filament radius with time is dependent on the balance of viscous, inertial, and capillary forces.[24] Therefore, it is not possible to apply Papageorgiou's solution (only valid without inertia), and the filament radius follows the following equation [55] at the latest states of the thinning process, right before breakup:

$$R_m = \frac{2X-1}{6}\frac{\sigma}{\eta}(t - t_b) \tag{4}$$

where $X$ is a constant that is determined experimentally at the last points of *R(t)* and $\sigma$ represents the surface tension of the fluid, which is assumed constant and independent of the applied electric field strength, following the same approach applied by Rubio *et al.* [56, 57] and Pelekasis *et al.* [58]. The value of the X constant is in average of $0.516 \pm 0.003$, close to the value of *X* determined by Eggers (*X = 0.5912*)[59], and independent on the electric field strength and the presence of the 2D nanoparticles.

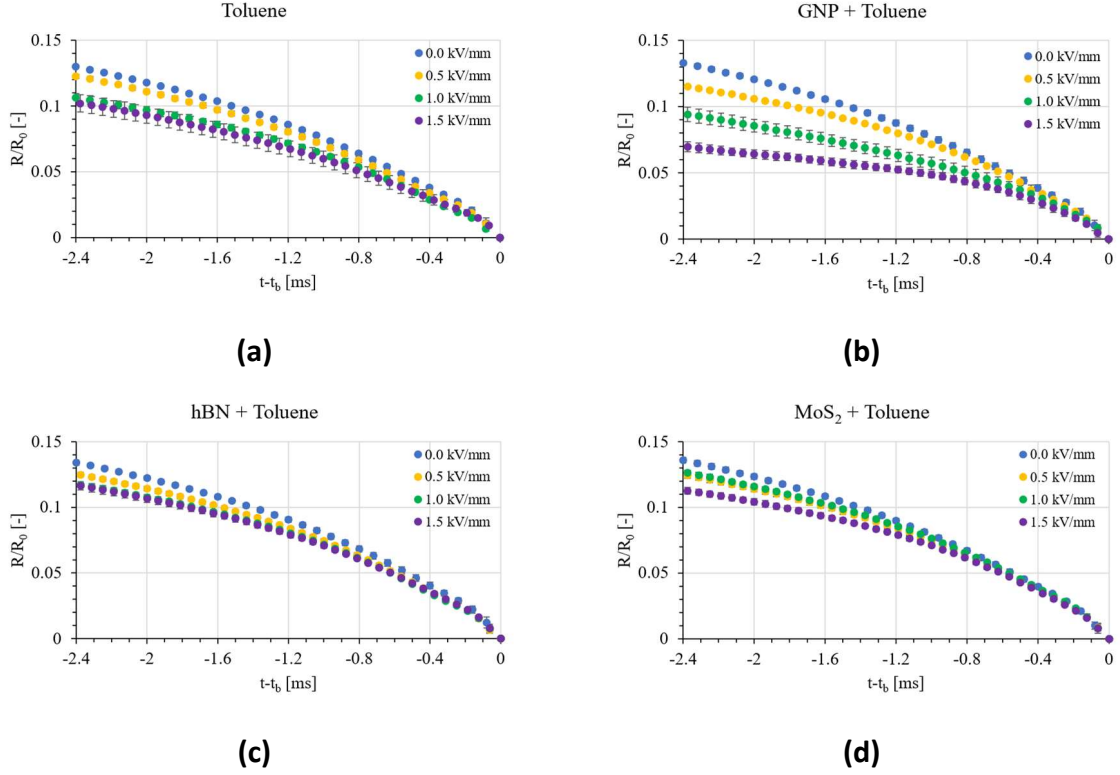

**Figure 8.** Time evolution of the minimum filament radius of **(a)** pure toluene, **(b)** toluene with GNP; **(c)** toluene withh hBN and **(d)** toluene with MoS$_2$ when the initial electric field strength is 0, 0.5, 1.0, and 1.5 kV/mm.

Pelekasis *et al.*[58] studied the stability of the liquid bridge from the equilibrium, , i.e. the liquid bridge underpinned fixed plates separated by a certain distance, under the presence of an external electric field. They defined the stability of the liquid bridge based on the ratio of the electrical conductivity between the internal fluid and the external fluid ($S = \kappa_{in}/\kappa_{ext}$), and the ratio between the dielectric constants between these two fluids ($\bar{\varepsilon} = \varepsilon_{in}/\varepsilon_{ext}$). **Table 2** shows the values of these ratios for Toluene (the carrier fluid) and air (the surrounding fluid).

**Table 2.** Electric properties of toluene and air.

| Fluid | $\kappa$ [S/m] | $\varepsilon$ [-] | $S$ [-] | $\bar{\varepsilon}$ [-] |
|---|---|---|---|---|
| **Toluene** | 8·10$^{-14}$ | 2.38 | 10 | 2.38 |
| **Air** | 8·10$^{-15}$ | 1.00 | | |

According to the analysis performed by Pelekasis *et al.* [58], when $S > \bar{\varepsilon}$, the tangential component of the electric stress promotes the stability of the liquid bridge; moreover, as the $(S-1)(\bar{\varepsilon}-1) > 0$, the normal component of the electric stress also promotes the stability of the liquid bridge. Therefore, it can be concluded that the Maxwell stresses help in stabilizing the filament during the thinning process in the CaBER experiments, decreasing in the slope of $R/R_0$ curves (**Figure 8**). Further, the electric conductivity of non-polar fluids depends on the electric field strength in which the electrical conductivity increases as the electric field strength increases.[46] So, the filament's lifetime of toluene increases, i.e., lower slope of $R/R_0$ curves shown in **Figure 8a**, since the tangential and normal components of the electric stresses will grow higher with increasing $E$.

**Figure 8Error! Reference source not found. -** *Figure 8*d show that the increase of the electric field strength slows down the thinning velocity, i.e., the local extension rate decreases, analogous to the previous observation on pure toluene. However, the presence of hBN and MoS$_2$ particles accelerate the thinning process for a specific electric field strength when compared to the GNP suspension. This significant variation in the slope of $R/R_0$ curves with the strength of electric field is related to the electrical conductivity of the particles used in this work, having the GNP particles the highest electrical conductivity, followed by MoS$_2$ particles and hBN particles have the lowest value. According to Rica *et al.*[60], the presence of particles in a suspension improves the dielectric constant of the fluid and the dielectric constant increases as the particle concentration increases. Further, Kharat *et al.* [61] observed that the addition of nanoparticles increases the electrical conductivity of the dispersion. Given these two assumptions and applying the same logic as Pelekasis *et al.* [58], the Maxwell stresses promote the stability of the liquid bridge, decreasing the rate of the filament thinning, being lower as the electrical conductivity of the nanoparticles increases. Moreover, the migration of nanoparticles follows the opposite direction of the electrical field, and it is more intense as the conductivity of the particles is higher (**Figure SI5** and **movie SI2**). Thus, as the particles migrate from bottom to top, they drag fluid towards the liquid filament, resulting in a longer filament life. Finally, it can also be observed an overlap of $R/R_0$ curves immediately before the filament breakup independently on the kind of 2D nanomaterial. Due to a very dilute particle concentration on the thinning zone, or eventually the total absence of particles, toluene controls the thinning process, similarly phenomenon observed by Rijo *et al.* [17]

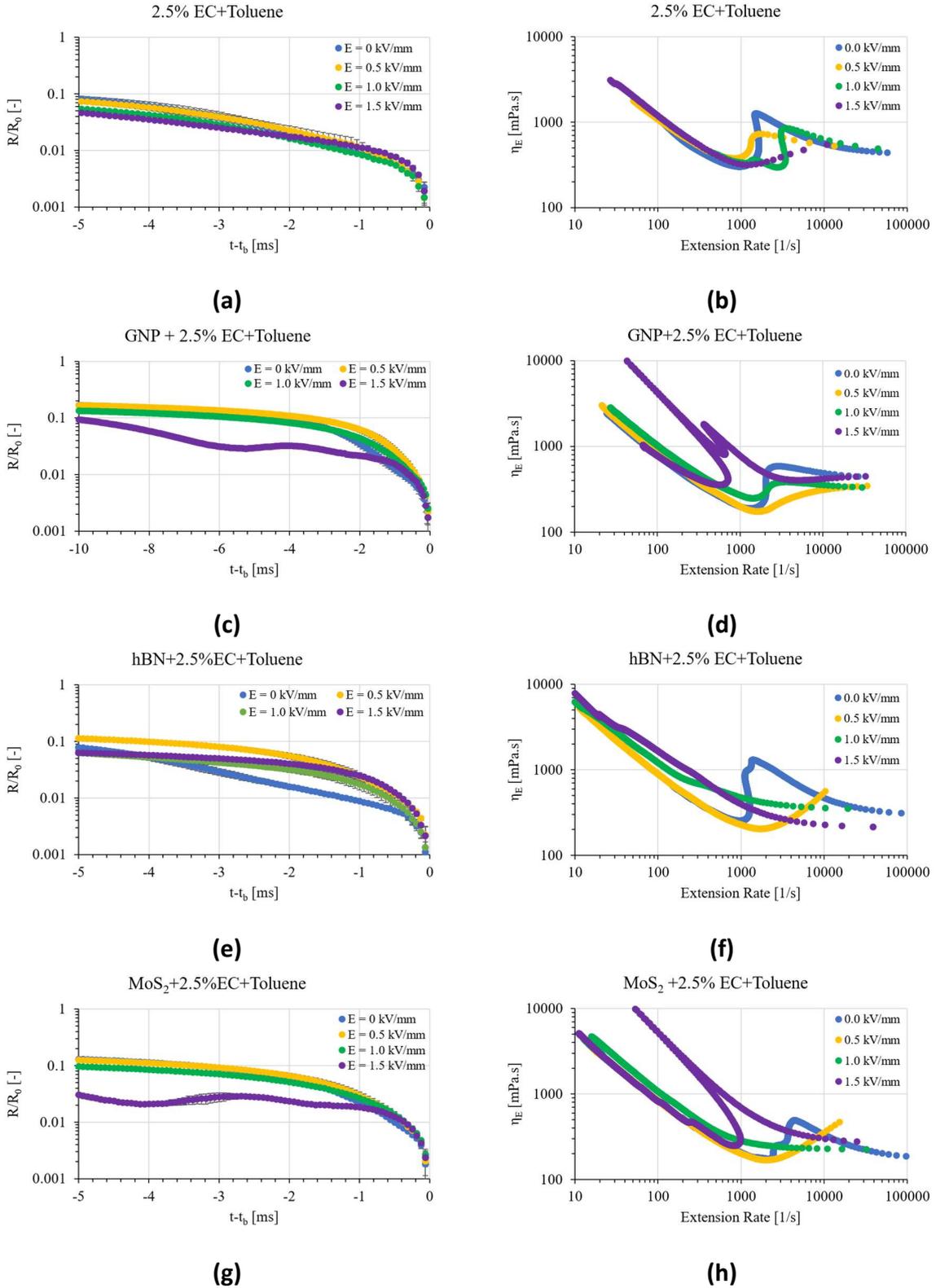

**Figure 9**. Time evolution of the minimum normalized radius of the filament (left) and the extensional viscosity curves (right) for **(a, b)** 2.5% w/v EC+Tol.; **(c, d)** GNP; **(e, f)** hBN and **(g, h)** $MoS_2$ dispersed in 2.5% w/v EC+Tol. when the initial electric field strength is 0, 0.5, 1.0, and 1.5 kV/mm.

**Figure 9a** shows that the addition of polymer vanishes the inertia effects when the liquid bridge is stretching. Furthermore, it is observed that the polymer solution has a longer lifetime compared to the toluene when the electric field is applied to the system. However, the presence of vortices seems to not affect the thinning rate as one would expect (**Figure 9**). Moreover, the presence of hydrodynamic motion inside of the filament is very similar to the hydrodynamic motion inside of Taylor cone reported in literature.[62] According to Barrero *et al.*[52], these movements are driven by the tangential electrical stress acting on the liquid-gas interface and the flow rate injected through the electrified needle. Here, there is not any flow rate injection during the CaBER experiments, so, the main driving forces are the tangential electrical stress. Further, the authors reported that an intense swirl can be observed when liquid with very small values of both electric conductivity and viscosity are used.[52]

Regarding the extensional viscosity curves, **Figure 9b** shows a decrease in the extensional viscosity until a critical extension rate is reached, upon which an increase of the extensional viscosity is observed, corresponding to the beginning of the uncoil of the polymeric chains. When the electric field strength is lower than 1.0 kV/mm, the increase of the $\eta_E$ is mild, reaching a maximum value before starting to decrease again until the filament breaks, which would mean that the polymer chains are fully stretched; however, when the E = 1.5 kV/mm, due to the stabilizing contribution of Maxwell stresses, the polymer chains do not have time enough to be fully stretched when E = 1.5 kV/mm.

When GNP particles are added to the 2.5% w/v ethyl cellulose polymer solution, **Figure 9c-d** shows that the application of electric fields above 1.0 kV/mm induce a slight filament thickening instead of thinning. That can only happen if the end-drops feed the liquid bridge with more fluid. A possible explanation is that the conductivity of the particles increases the stability of the liquid filament; then, the combination of vortex flow, induced by the presence of EC under the application of an external electric field, and the drag flow, induced probably by the particle migration from the bottom plate to the top (**Figure 10**), bring some fluid from the bottom end-drop to the liquid bridge. Thus, for E = 1.5 kV/mm, it was not possible to calculate the relaxation time because the presence of vortices and particle migration destroy the rheometric uniaxial extensional flow condition required for its calculation.

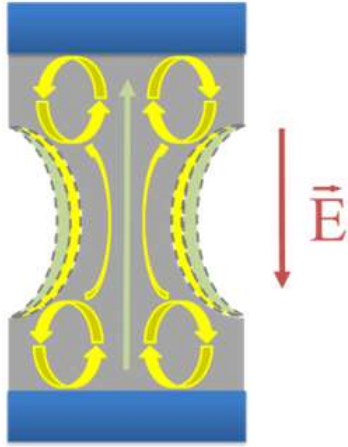

**Figure 10.** Sketch on the influence of vortex formation (yellow) and drag flow (light green) due to particle migration. If the conductive particles are present, the more stabilizing effect allows more fluid to enter the filament, both due to the vortex flow and the drag flow.

Below 1.5 kV/mm, the influence of these secondary flows does not have significant effect and the fluid relaxation time is determined in the region where a linear decay of the radius is present in a semi-log plot. This linear decay is fitted using the following equation:

$$\frac{R_{min}}{R_0} = \left(\frac{GR_0}{2\sigma}\right)^{1/3} \exp\left(-\frac{t}{3\lambda}\right) \qquad (5)$$

where $G$ is the shear modulus and $\lambda$ represents the relaxation time of the fluid. In the absence of electric field, the λ parameter corresponds to the relaxation time derived from the uncoiling of polymer chains. Here, the main forces present in the uniaxial elongational flow are the elastic force, and the surface tension. The gravity force and the inertial force are neglected since the Bond number is lower than 0.1, and Ohnesorge number is greater than 0.2077. When an electric field is present, two more forces come into play, which are the Maxwell force and the electrophoretic force. The presence of vortices and the particle migration invalidate the uniaxial elongational flow condition. Thus, in this case, the calculated λ parameter corresponds to an apparent relaxation time. **Table 3** shows the relaxation times (λ) of the 2.5% w/v EC solution without any particles remains constant when the electric field strength increases, even with the electrically induced vortices inside the fluid.

**Table 3.** Relaxation times (λ) for all suspensions having ethyl cellulose dissolved in toluene as carried fluid for several electric field strengths.

| EC [%] | Particles | E [kV/mm] | λ [ms] | Particles | E [kV/mm] | λ [ms] |
|---|---|---|---|---|---|---|
| 2.5 | No Particles | 0.0 | 0.451 ± 0.041 | hBN | 0.0 | 0.558 ± 0.031 |
| | | 0.5 | 0.474 ± 0.038 | | 0.5 | 0.240 ± 0.005 |
| | | 1.0 | 0.474 ± 0.061 | | 1.0 | – |
| | | 1.5 | 0.444 ± 0.022 | | 1.5 | – |
| | GNP | 0.0 | 0.341 ± 0.009 | $MoS_2$ | 0.0 | 0.171 ± 0.002 |
| | | 0.5 | 0.250 ± 0.008 | | 0.5 | 0.208 ± 0.003 |
| | | 1.0 | 0.303 ± 0.025 | | 1.0 | – |
| | | 1.5 | – | | 1.5 | – |
| 5 | No Particles | 0.0 | 5.19 ± 0.20 | hBN | 0.0 | 2.59 ± 0.14 |
| | | 0.5 | 18.8 ± 5.15 | | 0.5 | 1.06 ± 0.05 |
| | | 1.0 | – | | 1.0 | – |
| | | 1.5 | – | | 1.5 | – |
| | GNP | 0.0 | 1.22 ± 0.04 | $MoS_2$ | 0.0 | 1.79 ± 0.02 |
| | | 0.5 | 1.94 ± 0.03 | | 0.5 | 1.43 ± 0.03 |
| | | 1.0 | – | | 1.0 | – |
| | | 1.5 | – | | 1.5 | – |

Similar behavior is also observed for suspensions whose dispersed particles are $MoS_2$ (**Figure 9 g-h**), which conductivity is not as high as GNP but still high enough to make the liquid bridge stable and allow the fluid come in from the bottom end-drop due to the vortices and the particle migration. In the case of hBN (**Figure 9 e-f**), although the vortices are still present, the rheometric flow seems to not being affected by the addition of fluid from the end-drops; nevertheless, when the electric field is greater than 1.0 kV/mm, Maxwell stresses did not allow the full extension of the polymer chains.

Contrary to what was observed for the polymer solution of 2.5% of ethyl cellulose, when an electric field is applied to the polymer solution of 5% w/v of ethyl cellulose the filament thinning process was sensitively affected. It has been reported that the presence of an electric field induces polarization [63] and stretching of the polymer molecules [64]. This results in a change of the relaxation time and extensional viscosity depending on the intensity of the electric field (**Figure 11**). It becomes clear that the application of 0.5 kV/mm produces the optimal condition that allows the increase of the longest relaxation time of the solution, as can be seen from the λ values present in **Table 3**. Moreover, the viscosity of the polymer solution increases about 10 times when the polymer concentration is twice. According to Barrero *et al.*[52] and Gupta *et al.*[62], the viscosity damps or delays the formation of vortices when the electric field strength is below than the critical one. In this case, the critical electric field strength that induces the vortex formation is around 0.5 kV/mm. When the field exceeds 0.5 kV/mm, the relaxation time of the polymer solution cannot be calculated, since the vortices feed the filament and temporally prevent the thinning of the filament and the reduction of the applied extension rate at that time (**Figure 11**), which violates the criterion used to determine the relaxation time. Here, the relaxation time is determined when the extension viscosity increases as the extension rate remains constant or increases smoothly. **Figure 11** shows an increase in extensional viscosity followed by a decrease in extension rate, which means that the minimum radius of the filament remains practically constant over a short period of time.

The presence of 2D nanoparticles in the polymer solution with 5% w/v of ethyl cellulose affected in a similar way to what was observed for the 2.5% w/v EC. The larger the conductivity, the more important is the stability effect of the Maxwell stresses and the more intense is the drag flow associated with the particle migration. Due to the larger concentration of polymer, the critical intensity of the electric field to determine relaxation time was reduced to 1 kV/mm instead of 1.5 kV/mm.

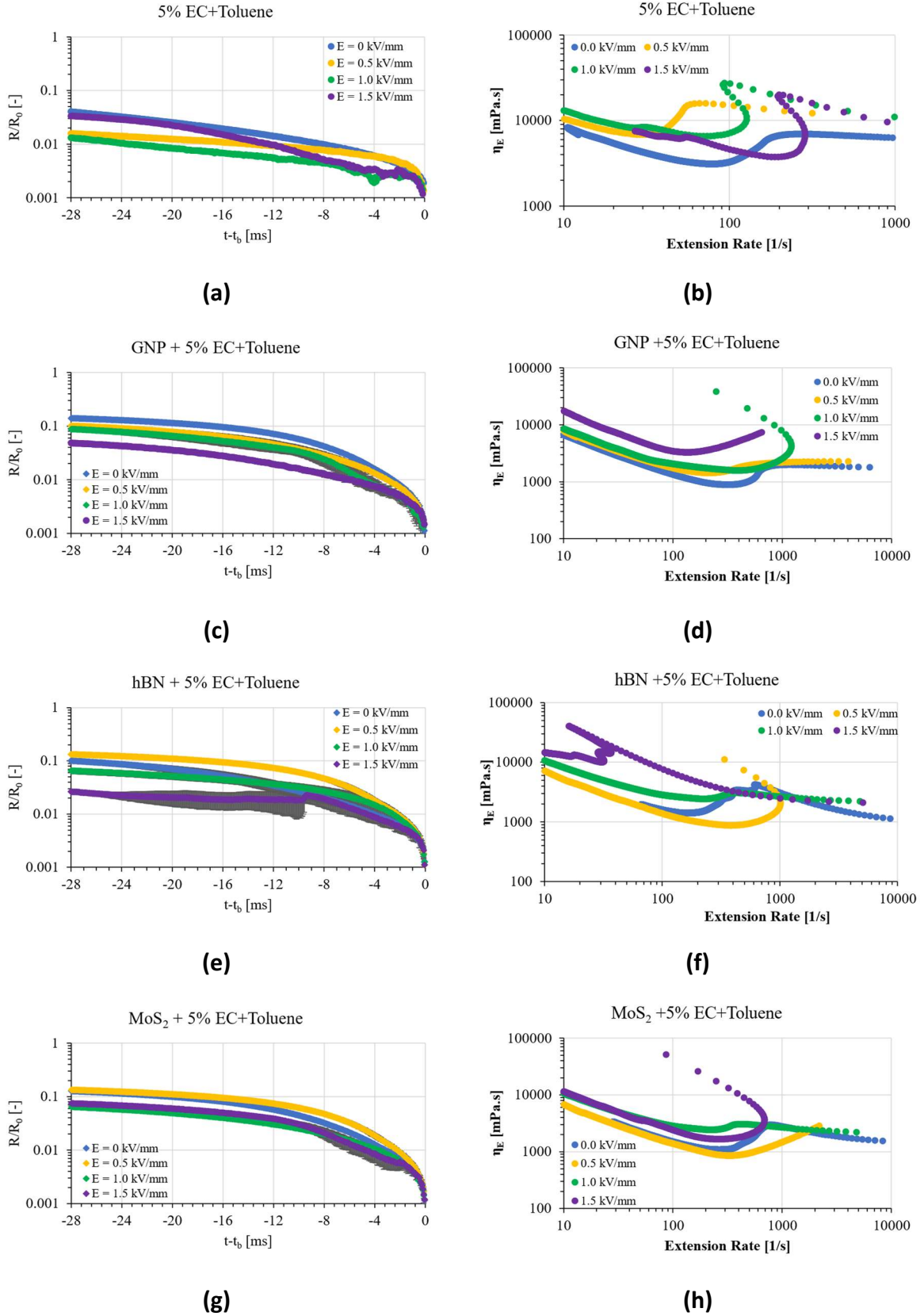

**Figure 11.** Time evolution of the minimum normalized radius of the filament (left) and the extensional viscosity curves (right) for **(a, b)** 5% w/v EC+Tol.; **(c, d)** GNP; **(e, f)** hBN and **(g, h)** $MoS_2$ dispersed in 5% w/v EC+Tol. when the initial electric field strength is 0, 0.5, 1.0, and 1.5 kV/mm.

## 4. Conclusion

This study represents pioneering work to understand how the effect of 2D nanoparticles affects the rheological properties of 2D-inks in the presence of an external electric field.

When an electric field is perpendicular to the flow direction, the 2D nanomaterials dispersed in toluene keep the Newtonian behavior and viscosity is almost constant and independent of the electric field strength. For 2D-inks formulations having Ethyl Cellulose, the shear stress slightly increases due to the presence of vortices due to an electrophoretic effect.

When electric field is aligned to the flow direction and no EC is present, the Maxwell stresses tend to stabilize the filament thinning process, helping the viscosity to counterbalance the action of surface tension; different 2D nanoparticles modifies the conductivity and the permittivity of the sample, modifying the intensity of the Maxwell stresses. The addition of EC induces the formation of vortices when the electric field is applied, as in shear experiments; these vortices also help in stabilizing the liquid bridges even more.

Based on the information here reported and the article review of Montanero and Gañan-Calvo [65], it is possible to estimate the range of droplet sizes in the cone-jet mode of electrospray for each suspension (**Figure 12**). When pure toluene is the carrier fluid, the more conductive the particle is, the smaller the droplet size. However, the presence of the EC allows to increase the conductivity of the fluids two orders of magnitude and, consequently, minimizing the impact of the conductivity of the particle. Assuming Newtonian behavior, the presence of EC would allow to reduce one order of magnitude the droplet size; however, the elasticity introduced into the system may affect this result and further research work is required in the future to be conclusive.

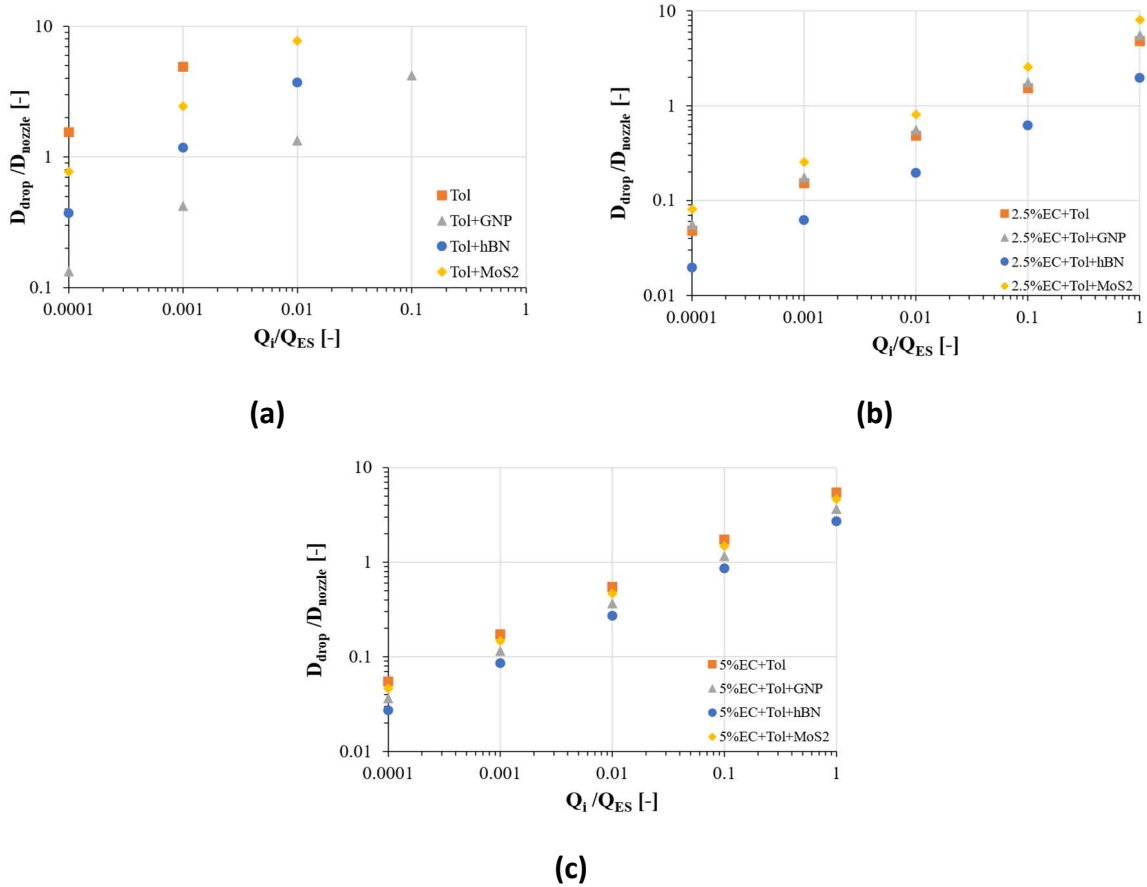

**Figure 12.** Normalized droplet size ($D_{drop}/D_{nozzle}$) vs $Q_i/Q_{ES}$ for the different carrier fluids **(a)** Toluene, **(b)** 2.5% w/v EC+Tol and **(c)** 5% w/v EC+Tol.

We understand that these pioneering results on the influence of different elements, such as the formulation (2D nanoparticles and EC) and the relative orientation between electric field and flow direction, on the Electrohydrodynamics of non-polar 2D inks will enlight the scientific community dealing with increasing the resolution of EHD printing techniques. Moreover, it sets the basis for validating numerical and theoretical studies that will be required to understand each force's strength, which are the building blocks behind the Electrohydrodynamics in the actual printing process.

**CRediT authorship contribution statement**
**Pedro C. Rijo:** Conceptualization, Methodology, Investigation, Formal analysis, Data curation, Writing - original draft

**Francisco J. Galindo-Rosales**: Conceptualization, Methodology, Formal analysis, Resources, Supervision, Writing - review & editing, Funding acquisition.

**Declaration of Competing Interest**

The authors declare that they have no known competing financial interests or personal relationships that could have appeared to influence the work reported in this paper.

**Data Availability.**

All data generated or analysed during this study are included in this published article [and its supplementary information files].

**Acknowledgements**

Authors acknowledge FEDER, FCT/MCTES (PIDDAC) and FCT for funding support under grants UI/BD/150886/2021, NORTE-01-0145-FEDER-000054, LA/P/0045/2020, UIDB/00532/2020 and UIDP/00532/2020. The authors also thank Graphenest for having graciously provided Graphene Nanoparticles (GNP), Prof. Cândido Duarte for helping in building a setup to measure the dielectric properties of fluids.